# GEOMETRIC ATTACK RESISTANT MULTILAYER IMAGE WATERMARKING SCHEME FOR PROVIDING HIGH SECURITY

**Dr.J.Veerappan[1]**
Department of Electronics and Communication Engineering,
Sethu Institute of Technology,
Viruthunagar-626106,

Tamil Nadu,India.

jveerappan@gmail.com

**G. Pitchammal[2]**
PhD Scholar , Anna university,
Chennai-600018,
Tamil nadu,India.
Pitchu.25@gmail.com

## Abstract

The main theme of this application is to provide an algorithm color image watermark to manage the attacks such as rotation, scaling and translation. In the existing watermarking algorithms, those exploited robust features are more or less related to the pixel position, so they cannot be more robust against the attacks. In order to solve this problem this application focus on certain parameters rather than the pixel position for watermarking. Two statistical features such as the histogram shape and the mean of Gaussian filtered low-frequency component of images are taken for this proposed application to make the watermarking algorithm robust to attacks and also AES technique is used to provide higher security.

## 1. INTRODUCTION

With the development of the Internet, more and more digital media products become available through different online services. The rapid growth of the multimedia services has created a potential demand for the protection of ownership since digital media is easily reproduced and manipulated. Digital watermarking has been introduced for solving such problems. One of the most prominent applications of watermarking is using robust and practical watermarking to protect image and video data [4]. In the past ten years, attacks against image watermarking systems have become more and more complicated with the development of watermarking techniques. In a desired watermarking system, the watermark should be robust to content-preserving attacks including geometric deformations and image processing operations. From the image watermarking point of view, geometric attacks mainly introduce synchronization errors between the encoder and decoder. The watermark is still present, but the detector is no longer able to extract it. Different from geometric attacks, the content-preserving image processing operations (such as addition of noises, common compression and filtering operations) do not introduce synchronization problems, but will reduce watermark energy.Most of the previous watermarking schemes have shown robustness against common image processing operations by embedding the watermark into the low-frequency component of images, such as the low-frequency subbands of discrete wavelet transform [2].

## 2. EXISTING HISTOGRAM BASED WATERMARKING ALGORITHMS

Some existing[6] histogram-based watermarking methods have been presented for the purposes of robust watermarking, fragile watermarking and reversible watermarking. The well-known patchwork watermarking methods inserted a message by supposing that two sets of randomly selected pixels are Gaussian distributed with zero mean.[7] The watermark sequence was embedded by shifting the mean values between groups of two sets of pixels. Propose a reversible watermarking scheme based on interpolation which features very low distortion and relatively large capacity. Different from previous watermarking schemes, utilize an interpolation technique to generate residual values and expand them by addition to embed bits. The strategy is efficient since interpolation values good at decorrelating pixels and additive expansion is free of expensive overhead information [2]

## 3.PROPOSED WATERMARKING ALGORITHM

In the existing methods, the watermark is difficult to survive under RBAs because the exploited robust features are more or less related to the pixel position. Thus, a possible way to cope with RBAs is to embed the watermark into those feature representations independent of the pixel position. Towards this direction, we use the scaling invariance of shape of an image's histogram[15] and the property of the histogram to be independent of the pixel position against various geometric attacks, which is implemented by representing the histogram shape as the ratios of population between groups of two neighboring bins and then modify the ratios to carry a key-based encrypted pseudo-random noise (PN) sequence. Considering the interpolation errors and common image processing operations, the two features are extracted from the Gaussian filtered low-frequency component. The Advanced Encryption Standard (AES) [16]computer security standard is a symmetric block cipher that encrypts and decrypts 128-bit blocks of data. Standard key lengths of 128, 192, and 256 bits may be used. The algorithm consists of four stages that make up a round which is iterated 10 times for a 128-bit length key, 12 times for a 192-bit key, and 14 times for a 256-bit key.

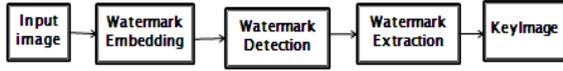

Fig.1 Watermarking Framework.

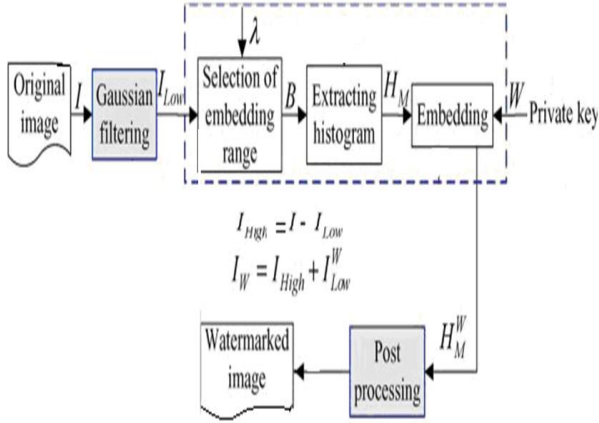

Fig.2 Watermark embedding framework

A. Watermark Insertion

As illustrated in Fig. 2, the watermark insertion consists of three main steps: Gaussian filtering, histogram-based embedding and post processing.

1) Gaussian Filtering: The input image (I) is filtered with a Gaussian kernel low pass filter for removing the high-frequency Information $I_{High}$. In the low pass filtering operation can be represented as the convolution of the Gaussian function $G(x,y,\sigma)$ and an image $I(x,y)$,

$$I_{Low}(x,y) = G(x,y,\sigma) * I(x,y) \qquad (2)$$

In 2-D case, an isotropic (i.e., circularly symmetric) Gaussian function has the form

$$G(x,y,\sigma) = \frac{1}{2\pi\sigma^2} \exp^{-\frac{x^2+y^2}{2\sigma^2}}$$

(3)

where $\sigma$ is the standard deviation of the distribution. Because $G(x,y,\sigma)$ is isotropic and circular in shape, the resulting response $I_{Low}(x,y)$ is invariant to rotation, which is beneficial for obtaining a geometric distortion-resilient watermark.

2) Histogram-Based Embedding: The histogram ($H_M$) is extracted from the filtered image ($I_{low}$) by referring to its mean value in such a way that the watermark is immune to the operation of scaling the value of all pixels, and the encrypted watermark recovery process can avoid exhaustive search. The encrypted watermark, denoted by W= {$w_i$|i=1,….,$L_w$} is a key-based PN sequence. The private key is shared with the detector during the decision-making for presence of the watermark as used in [11].The average value of the low-frequency component $I_{Low}$ of an image is calculated as $\bar{A}$. By referring to $\bar{A}$, an embedding range denoted by B= [(1 - λ) $\bar{A}$, [(1 + λ)$\bar{A}$] is computed to generate the histogram with L equal-sized bins of width M , denoted by H={$h_M(i)$|i=1,….,L}. L should not be less than $2L_w$ in order to embed all bits.

Let Bin_1 and Bin_2 be two consecutive bins in the extracted histogram. Suppose their population is a and b, respectively. The embedding rules are formulated as

a / b ≥ T,   if   w(i) = 1

b /a ≤ T,   if   w(i) = 0            (4)

where T is a threshold controlling the number of modified samples.

Consider[9] the case that w(i) is bit value "1." If a / b ≥ T, no operation is needed. Otherwise, the number of pixels in two neighboring bins, a and b , will be reassigned until satisfying the condition $a_1$ / $b_1$ ≥ T . When w(i) is "0," the procedure is similar. If w(i) is "1" and a / b < T, $I_1$ randomly selected samples from Bin_2 will be modified to Bin_1, achieving $a_1$ / $b_1$ ≥ T . If w(i) is "0" and"1" , randomly selected pixels from Bin_1 will be moved to Bin_2, satisfying $b_0$/ $a_0$≥ T .

$f'_1(i) = f_1(i) + M, 1 \leq I \leq I$

$f'_2(j) = f_2(j) - M, 1 \leq j \leq I_1$ (5)

Where M is the bin width, $f_1(i)$ is the $i^{th}$ selected pixel in Bin_1, and $f_2(j)$ denotes the $j^{th}$ selected pixel in Bin_2. The modified pixels $f'_1(i)$ and $f'_2(j)$ belong to Bin_2 and Bin_1, respectively. $I_0$ and $I_1$ can be computed by the following expressions:

$I_0 = T \cdot b - a / 1 + T$, $I_1 = T \cdot a - b / 1 + T$

where $a_1 = a + I_1, b_1 = b - I_1, a_0 = a - I_0$, and $b_0 = b + I_0$. (6)

3) Post Processing: According to the property of the Gaussian kernel filter, we know the extracted watermark energy $G(I_w - I)$ is different from the embedded watermark energy $I^w_{Low} - I_{Low}$. The difference reflects a loss of the watermark energy caused by the Gaussian filtering in the extraction. Considering the effect, we introduce a corresponding post processing step in the watermark embedding phase by computing the maximum difference of the pixels between $G(I_w - I)$ and $I^w_{Low} - I_{Low}$, formulated (4) as

$dGau = \max(|G(I_w - I) - (I^w_{Low} - I_{Low})|) - M$. (7)

Then, the histogram $H^w_M$ is further modified to compensate the loss of the watermark energy as follows:

$f(i) = (k-1)M + \mu$, if $i < (k-1)M + \mu$
$f(i) = kM - \mu$, if $i > kM - \mu$ (8)

B. Watermark Recovery:

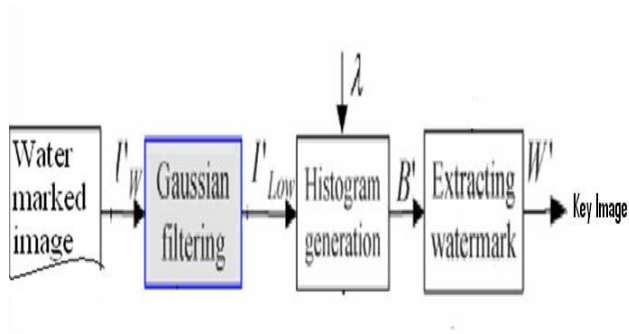

Fig:3 Watermark extraction framework

First, we use σ (the same as in the embedding) for the Gaussian filtering to start the mean-based search process.

1. Compute the histogram of $I'_w$ from the range $B = [(1-\lambda)\bar{A}', [(1+\lambda)\bar{A}']$ with L equal-sized bins as in the process of watermark embedding.

2. Divide the histogram bins as groups, two neighboring bins as a group. Suppose that the population in two consecutive bins, are a' and b', respectively. By computing the ratio between a' and b', one inserted bit is extracted in reference to the following equation,

$w'(i) = 1$, if $a'' / b \geq 1$

$w'(i) = 0$, otherwise. (9)

The process is repeated until all bits are extracted.

## 4. PERFORMANCE ANALYSIS

In this section, we evaluate the performance of the proposed watermarking algorithm in terms of peak signal-to-noise ratio (PSNR) computation, embedding capacity, and computational cost in the extraction, false positive probability and possible key-estimation attack.

A. PSNR Computation:

In the watermarking algorithm, the high-frequency component is unchanged. That is, the watermark is only inserted in the low-frequency component to generate watermarked image

PSNR is defined as:

$PSNR = 20\log_{10}(MAX^2/MSE)$ (10)

MAX-maximum pixel value of the frame

MSE-mean squared error

The above conclusion is very useful for reducing the computational cost in the embedding since we use to control the embedding distortion.

B. Capacity:

Suppose that the mean of the Gaussian filtered image is Q and the parameter P is applied to compute the embedding region. The embedding capacity $L_w$ of the proposed algorithm can be expressed as

$L_w = 2P \cdot Q / M \cdot G$ (11)

where M denotes the bin width, and G is the number of the bins in a group designed to embed one bit. In this paper, G is 2 (two neighboring bins as a group). For images of D-bit depth, the maximum embedding capacity of the watermarking algorithm is mathematically calculated as $2^D/M \cdot G$ which can achieve 128 bits for the 8-bit images in case of M=1. In case that the parameters are given as Q=128(half of 256 gray levels), P=0.6, M=2 and G=2, the watermark embedding capacity is estimated as $L_w = (2*0.6*128)/(2*2) = 40$.

Considering the fact that the mean values for some images are far less than 128, the effective embedding capacity of the proposed algorithm is from 20 to 30 bits.

C. Watermark Security Discussion:

We can see that the length of the exploited PN sequence has an important role on the security of the watermarking scheme. In the proposed framework, the encrypted [16]information is embedded into the histogram bins of the low-frequency component. With consideration of the robustness, the number of the bins is limited (refer to Section V-B). Thus, the exploited PN sequence is limited in length, and cannot provide a sufficient randomization. Once the watermarking algorithm is public, an attacker may use such a lack of randomization to mount an attack to estimate the parameter P by matching the extracted bits with the randomly generated PN sequence, and then modify the bins to remove the watermark.

## 5.EXPERIMENTAL RESULTS

In this section, we first investigate the effect of the Gaussian filter-based preprocessing on the watermark extraction for the design of the post processing operation in Section III. Then, the watermark imperceptibility and robustness are evaluated by using 50 different 512 images as example images, respectively. The comparision results for different watermarking techniques with various λ is shown in fig4.The example images include 48 standard grayscale test images and Benchmark image Leon. In the experiments, a 16-bit of PN sequence was embedded into the 50 histogram bins. The comparision results for different watermarking techniques with different images is shown in table1.

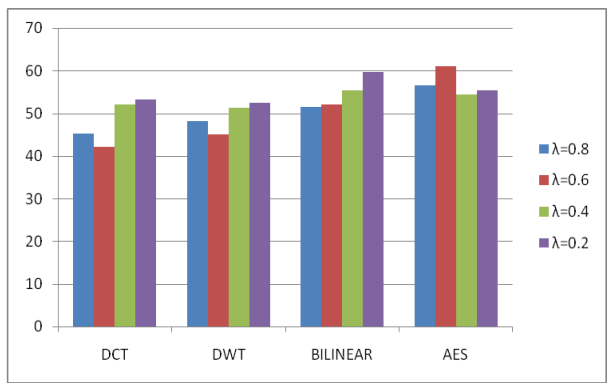

Fig:4 Different watermarking Techniques for various λ

Table1:Comparision of Results.

| Input Image | Key Image | Different watermarking techniques(PSNR) | | | | |
|---|---|---|---|---|---|---|
| | | AES | DCT | DWT | LZW | Bilinear |
| Natural scene | Barb | 57.26% | 51.13% | 53.21% | 55.61% | 58.63% |
| Lena | Number | 61.23% | 50.16% | 52.36% | 54.26% | 60.23% |
| Pepper | B.E | 58.4% | 52.36% | 51.23% | 51.53% | 58.23% |
| Flower | CSE | 52.13% | 42.45% | 48.36% | 49.21% | 50.12% |
| Tiger | Barb | 55.23% | 46.28% | 51.45% | 51.12% | 54.12% |

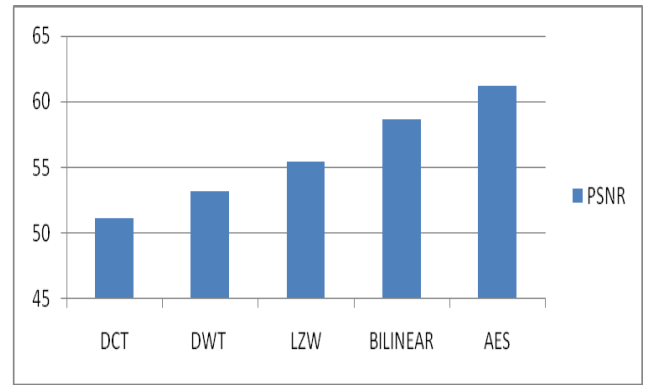

Fig:5 Different watermarking Techniques for sample images

A.Effect of the Gaussian Filtering

As discussed in Section III-A, a post processing step is designed to compensate the loss of the watermark energy due to Gaussian filtering in the extraction. We can see that the maximum absolute errors for all the 50 test images are between 0.2 to 0.75. Thus, in the post processing phase the parameter μ is set equal to 0.75 for tolerating the effect.

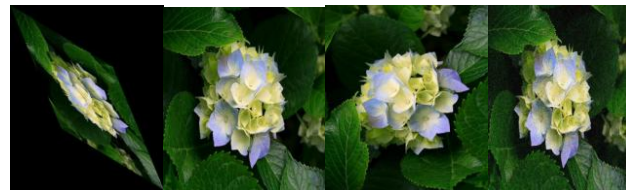

Fig.6 Transformed Images. (a) XY Shearing (b) Scaling

(c) Rotation (d) Gaussian Noise

B. Watermark Imperceptibility

Taking Leon as example, we plot the watermarked images and the amplitude of the watermark in Fig. a. We can see that the watermarked images are perceptibly similar to the original images. The maximal pixel distortion in the spatial domain due to the watermark is within 3 gray levels. In Fig. b, we investigate the effect of the watermark on the mean. The mean values of these images remain almost unchanged after watermarking (the mean errors are between 0.15 and 0.25). The reason is that the pixels are marked by adding or reducing the bin width with the same probability in reference to(2). It ensures the recovery of the watermark from a marked image. The PSNR values of the 50 watermarked images are between 44 and 58 dB. The above analysis shows that the embedding distortion can be controlled by using the threshold T. In our experiments, only part of the samples is modified with a low amplitude, which is beneficial to generating the high quality watermarked images.

## 6. CONCLUSION

In both theoretical analysis and experimental method, in this paper we present a robust image watermarking algorithm against various attacks including challenging cropping operations and RBAs by using the property of the histogram shape to be independent of the pixel position, mathematically invariant to the scaling, statistically resistant to cropping. A key-based PN sequence is successfully inserted by modifying the histogram shape, which is computed from the low-frequency component of Gaussian filtered images by referring to the mean. The encrypted watermark is generated by AES algorithm. The watermark can be detected without knowledge of original images by sharing the exploited private key in the detector.

## 7. FUTURE ENHANCEMENT

In our future research, one consideration is to enhance the security of the watermarking scheme by introducing the color information of color images, video and audio signal or seeking new ways to watermark the histogram shape so that a longer PN sequence can be embedded.

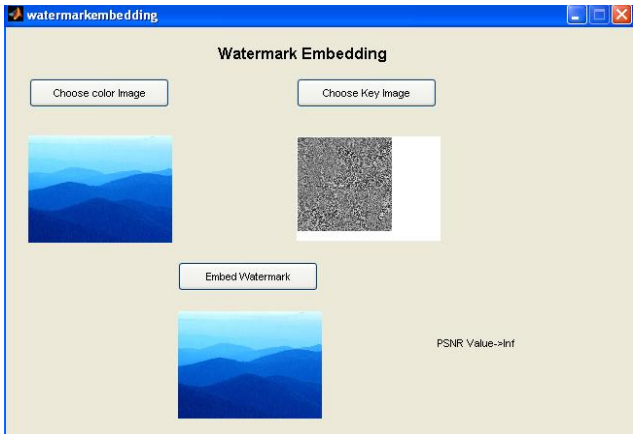